\documentclass[12pt,a4paper]{article}
\usepackage{graphicx,amssymb}
\usepackage{graphicx, array, epsfig, subfigure}
\usepackage{multirow}
\usepackage{amsmath}
\usepackage{graphicx}
\usepackage{adjustbox}
\usepackage{multirow}
\usepackage{booktabs,caption}
\usepackage[flushleft]{threeparttable}
\usepackage{float}
\usepackage{subfigure}  
\newcommand{\be}{\begin{equation}}
\newcommand{\ee}{\end{equation}}
\newcommand{\bea}{\begin{eqnarray}}
\newcommand{\eea}{\end{eqnarray}}

\begin{document}
\begin{center}
{\bf Gravitational Wave Emissions from First-Order Phase Transitions with Two Component FIMP Dark Matter}\\
\vspace{1cm}
{{\bf Madhurima Pandey} \footnote{email: madhurima.pandey@saha.ac.in},
{\bf Avik Paul} \footnote{email: avik.paul@saha.ac.in}}\\
{\normalsize \it Astroparticle Physics and Cosmology Division,}\\
{\normalsize \it Saha Institute of Nuclear Physics, HBNI} \\
{\normalsize \it 1/AF Bidhannagar, Kolkata 700064, India}\\
\vspace{1cm}
\end{center}
\begin{abstract}
We explore the emissions of the Gravitational Waves (GWs) from a strong first-order ekectroweak phase transition. To this
end, a dark matter model has been investigated in Feebly Interacting Massive
Particle (FIMP) scenario, where the
dark matter particles are produced through ``freeze-in'' mechanism in the
early Universe and due to their very small
couplings they could not attain thermal and chemical equilibrium with
the Universe's thermal plasma. In this context,
we extend scalar sector of Standard Model of particle physics by
two additional scalar singlets whose stability
is protected by an unbroken discrete $Z_2 \times Z'_2$ symmetry and they
are assumed to develop no VEV after
spontaneous symmetry breaking. We study the first-order phase transition
within the framework of this present model.
We have done both analytical and numerical computations to calculate
the consequent production of GWs and then the
detectabilities of such GWs have been investigated  at the future space
based detectors such as LISA, BBO, ALIA, DECIGO, aLIGO and aLIGO+ etc. We also find that dark matter self coupling has a considerable influence on the GW
production in the present scenario.
\end{abstract}
\newpage
\section{Introduction}
In 2015 the Laser Interferometer Gravitational Wave Observatory (LIGO) 
collaboration \cite{ligo} had detected Gravitational 
Waves (GW150914) reaching the Earth from the distant violent astrophysical 
phenomenon involving inward spiralling and eventual merger of a pair of 
massive black holes and subsequent ringdown after the merger
opens up new vistas to Gravitational Wave astronomy. This has been followed
by the detections of a number of such other GWs and the study of these 
GWs helps to explore high energy cosmic phenomena such as neutron 
stars, pulsars, black holes etc. 
Other than the likes 
detected by LIGO where the GWs are generally originated from the collision 
of black holes and/or neutron stars, primordial GWs can be 
produced via inflationary quantum fluctuations \cite{Starobinsky:1979ty}, 
topological defects of the domain walls and cosmic strings \cite{Vilenkin:2000}, first-order phase transitions in 
the early Universe \cite{Witten:1984rs, {Hogan:1986qda}} etc. 
Study of these GWs could be very useful to understand the cosmic evolution
and the processes in the early Universe. 
In this work, we address the case where the GWs can be produced from 
first-order electroweak phase transition mediated by a viable dark 
matter candidate. Similar issues have been addressed in earlier works \cite{Kozaczuk:2014kva}-\cite{Huang:2017rzf}
but here we consider a feebly interacting massive particle or FIMP 
to be the dark matter candidate and it is a two component dark matter
instead of the usual approach with a one component dark matter.        
The simple extension of Standard Model (SM) of particle physics (SU(2)$_{\rm L}$ 
$\times$ U(1)$_{\rm Y}$ electroweak symmetry) with additional particles 
(to accommodate the two component particle dark matter candidate) may 
lead the  elctroweak phase transition to be a first-order phase 
transition instead of a just a smooth crossover. 

The electroweak phase transition (EWPT) 
followed by the spontaneous symmetry breaking 
(SSB) of 
electroweak symmetry (SU(2)$_{\rm L} \times $ U(1)$_{\rm Y}$) as formulated in SM is a smooth transition 
rather than a first-order one. In this paper, GWs from a first-order electroweak phase transition is addressed in the 
framework of a proposed dark matter (DM) model. Though the Standard Model of particle physics 
can describe the descriptions of several experimental observations, it may not be considered as a
fundamental theory as it fails to describe the dark matter phenomena and the baryon asymmetry of the Universe (BAU). In 
order to explain the latter a strong first-order electroweak phase transition (SFOEWPT) is required which can provide a 
non-equilibrium environment if the electroweak baryogenesis mechanism 
\cite{Kuzmin:1985}-\cite{Morrissey:2012} generates the BAU. There are many models 
and theories where the extension of the SM has been studied to address the EWPT and the DM phenomena simultaneously 
\cite{Kozaczuk:2014kva}-\cite{Huang:2017rzf}. Different types of phase transitions (PT) namely one-step \cite{one1}-\cite{one5} and 
two-step (or more step) \cite{two}-\cite{two9} PT , may lead to a 
SFOEWPT. For one-step PT only initial and final phases exist whereas the temperature of the Universe drops down to 
the barrier between the electroweak symmetry and the broken phases appear from loop corrections of the potential. 
On the other hand for two-step (or multi step) PT an intermediate metastable state arises between the initial and the 
final phases and the barrier between two minima appears at the tree level of the potential. As a result the transition 
may happen as a first-order phase transition through a tunneling process and it may proceed through electroweak 
bubble nucleation \cite{bubble1}-\cite{bubble3}, which can expand, collide and 
coalesce and eventually the Universe turns into the electroweak 
broken phase. Initially all possible sizes of bubbles having different surface tension and pressure are taken into account. 
The bubble with the size which is large enough to avoid any kind of collapses are claimed as a critical bubble. 
The bubbles having size smaller than the critical one tend to collapse whereas the bubbles having size larger than the 
critical bubble tend to expand. Due to the pressure difference between the true and false vacua, nucleations of 
electroweak bubbles occur. These bubbles could not sustain their spherical symmetry which generates 
phase transitions and eventually these phase transitions may produce the Gravitational Waves. The GW production 
mechanisms through the bubble nucleation and bubble collision have been 
addressed in literatures \cite{Kosowsky:1992a}-\cite{Caprini:2008}. The three mechanisms based on which the GW production 
arises mainly are bubble collisions and shock waves \cite{Kosowsky:1992a}-\cite{Caprini:2008}, sound waves \cite{Hindmarsh:2014}-\cite{Hindmarsh:2015} and 
magnetohydrodynamic turbulance 
\cite{Caprini:2006}-\cite{Caprini:2009} in the plasma.

As mentioned earlier, the particle nature of DM cannot be effectively explained by SM. 
In this work we address the issues of SFOEWPT as well as the particle dark matter candidate by adopting a minimally 
extended SM where two real scalar singlets are added to the SM. This is following well motivated Feebly Interacting 
Massive Particle (FIMP) scenario \cite{Yaguna:2011qn,Molinaro:2014lfa,zakeri,{amitanirban}}, which 
is an alternate to the WIMP (Weakly Interacting Massive Particle) mechanism \cite{Jungman:1995df}-\cite{Biswas:2013nn}. FIMPs can be identified by their feeble 
interactions with SM particles in the early Universe. As a result FIMP particles could not attain thermal equilibrium 
with the thermal bath of the Universe. The production of FIMP particles can be initiated by annihilation or decays of 
the SM particles in the thermal bath and in contrast to WIMP scenario (in thermal equilibrium) they are unable to 
undergo annihilation interactions 
among themselves due to their extremely negligible initial abundances. Hence they could not reach thermal and chemical 
equilibrim with the Universe's thermal plasma. However the number densities 
of FIMPs increases slowly and gradually due to very small couplings with other particles in the plasma. 

In this work, the scalar sector of SM is extended by adding two real scalars $S_1$ and $S_2$ and both of these scalars 
are singlets under the SM gauge group SU(2)$_{\rm L} \times $ U(1)$_{\rm Y}$. We impose a discrete $Z_2 \times Z'_2$ 
symmetry on 
the two scalars to prevent their interactions with the SM particles or their decays and they do not generate any vacuum 
expectation value (VEV) at SSB. 
The Higgs portal is the only way through which these scalars can interact with the SM particles. The bounds and the 
constraints on the parameters of our proposed dark matter model from both theoretical and experimental aspects are 
elaborately dicussed by one of the authors in an earlier work \cite{madhu}. The two component FIMP dark matter in this model 
also satisfies the upper limit of 
the dark matter self interacting cross-sections obtained from the observations of 72 colliding galaxy clusters 
\cite{Harvey:2015hha} and this is discussed in Ref \cite{madhu}. Though the relic density calculations of the 
proposed dark matter model can be done by considering the Higgs portal, however the couplings of the dark matter 
self interactions dominate over the Higgs portal couplings in the case of generating SFOEWPT as the couplings of DM 
with the SM particles are very negligible $\sim 10^{-9} - 10^{-11}$. Our purpose is to demonstrate the viability 
of the present extended SM to induce the SFOEWPT and at the same time yields a viable dark matter candidate. 
We have chosen three benchmark points from the allowed parameter space for the analytical and numerical 
calculations of the GW production due to SFOEWPT in the present model. We have checked that the three chosen 
benchmark points can satisfy the criterion $\dfrac{v_c}{T_c} \gtrsim 1$ 
\cite{sha1}-\cite{sha3}, where $v_c$ indicates the Higgs VEV at the 
critical temperaure $T_c$, which is required to produce SFOEWPT and hence the GWs are generated. We also address 
the detectability of such GWs by calculating the GW frequencies and intensities for the present formalism and comparing them with the estimated 
sensitivity predictions of the future space borne detector such as Big Bang Observer (BBO) \cite{Harry:2006}, 
Laser Interferometer Space Antenna(LISA)~\cite{Caprini:2016}, 
Advanced Laser Interferometer Antenna (ALIA) \cite{Gong:2015}, DECi-hertz Interferometer Gravitational wave 
Observatory (DECIGO) \cite{Seto:2001}
and ground-based detector such as advanced Laser Interferometer Gravitational-Wave Observatory (aLIGO) \cite{Harry:2010}.

The remaining of this present paper is arranged as follows. In 
Section 2 we study the two component dark matter model where the SM is extended by adding two real singlet scalars. 
Section 3 addresses the capability of a strong first-order phase transition to generate the Gravitational Wave signals and 
it contains two subsections. In Subsection 3.1 we discuss about the finite temperature effective potential whereas the 
production mechanism of the Gravitational Waves are furnished in Subsection 3.2. In Section 4 we explore the GW calculations 
and its discovery prospects. Finally, in Section 5 we present a summary and conclude our work.

\section{Two component FIMP dark matter model}
In this work, a two component FIMP dark matter model is adopted, where the dark 
matter candidates are produced through freeze-in mechanism. The model is a 
renormalisable extension of the Standard Model of particle physics with two singlet scalar fields 
$S_1$ and $S_2$. Therefore, these two real scalars ($S_1$ and $S_2$) compose 
the dark sector in the present scenario. The scalars $S_1$ and $S_2$ are singlets under the 
SM gauge group (${\rm SU(2)}_{\rm L} \times {\rm U(1)}_{\rm Y}$). We impose 
a discrete $Z_2 \times Z'_2$ symmetry on the real scalar fields to ensure its 
stability. After spontaneous symmetry breaking 
(SSB), these scalars do not generate any vacuum expectation value (VEV), 
which prevents mixing between the real scalars and the SM scalar. The DM candidates can interact 
only through the Higgs portal with the SM sector. 
The Lagrangian of the model is given as 
\bea
{\cal L} = {\cal L}_{\rm SM} + {\cal L}_{\rm DM} + {\cal L}_{\rm int}\,\, ,
\label{model1}
\eea
where ${\cal L}_{\rm SM}$ represents the Lagrangian for the SM particles 
having the usual kinetic term as well as the quadratic and quartic terms 
for the Higgs doublet $H$. ${\cal L}_{\rm DM}$ stands for the dark sector 
Lagrangian including two real singlet scalars, which can be written as 
\bea
{\cal L}_{\rm DM} = {\cal L}_{S_2} + {\cal L}_{S_3}\,\, ,
\label{model2}
\eea
with 
\bea
{\cal L}_{S_i} = \frac{1}{2} (\partial_{\mu} S_i)(\partial^{\mu} S_i) -
\frac{{ \mu}_{S_i}^2}{2} S_i^2 - \frac{\lambda_{S_i}}{4} S_i^4\,\, ,
\label{model3}
\eea
where $i = 1,2$. The interaction Lagrangian (${\cal L}_{\rm int}$) for all 
possible mutual interaction terms among the scalar fields $H, S_1$ and $S_2$ is expressed as 
\bea
{\cal L}_{\rm int} = -V' (H, S_2, S_3)\,\, ,
\label{model4}
\eea
where $V' (H, S_2, S_3)$ is given as 
\bea
V^{\prime}(H,\,S_2,\,S_3) ={\lambda_{H S_1}} H^\dagger H\, S_1^2
+ {\lambda_{HS_2}} H^\dagger H\,S_2^2
+ \lambda_{S_1 S_2} S_1^2\,S_2^2 \,\, .
\label{model5}
\eea
Let us consider the renormalisable potential term of the scalar sector of the 
two component dark matter model in FIMP scenario ($V$) as 
\bea
V &=& \mu_{H}^2\,H^\dagger H + \lambda_{H}\,(H^\dagger H)^2
+ \frac {\mu_{S_1}^2} {2} S_1^2 + \frac {\lambda_{S_1}}{4} S_1^4
+ \frac{ { \mu}_{S_2}^2}{2} S_2^2 + \frac{\lambda_{S_2}}{4} S_2^4 \nonumber \\
&& + {\lambda_{H S_1}} H^\dagger H\, S_1^2
+ {\lambda_{HS_2}} H^\dagger H\,S_2^2
+ \lambda_{S_1 S_2} S_1^2\,S_2^2 \,.
\label{model6}
\eea
After SSB, since $S_1$ and $S_2$ do not acquire any VEV, we have 
\be
H = \frac {1} {\sqrt {2}} \left ( \begin{array}{c} G^+ \\ (v + h + i G_0)
\end{array}
\right )\,\, , S_1 = s_1 + 0\,\, , S_2 = s_2 + 0 \,\, ,
\label{model7}
\ee
where $G^+$ and $G_0$ indicate the charged and neutral Goldstone bosons 
respectively and $\langle H \rangle = v$, is the Higgs VEV. 
The gauge bosons ($W^{\pm}$, Z) abdorb the Goldstone bosons 
($G^+, G_0$) after SSB \cite{Banik:2014eda}. 

As a result, after SSB the scalar potential ($V$) takes the form 
\bea
V&=& \frac{\mu_H^2} {2} (v + h)^2 + \frac{\lambda_H} {4} (v + h)^4 +
\frac{\mu_{S_1}^2} {2} s_1^2 + \nonumber \\
&& \frac{\lambda_{S_1}} {4} s_1^4 +
\frac{\mu_{S_2}^2} {2} s_2^2 + \frac{\lambda_{S_2}} {4} s_2^4 + \nonumber\\
&& \frac{\lambda_{HS_1}} {2} (v + h)^2 s_1^2 +
\frac{\lambda_{HS_2}}{2} (v + h)^2 s_2^2 +
\lambda_{S_1 S_2} s_1^2 s_2^2 \,\, .
\label{model8}
\eea
By using the minimisation condition 
\be
\left (\frac{\partial V}{\partial h}\right),\,
\left(\frac{\partial V}{\partial s_1}\right),\,
\left ( \frac {\partial V} {\partial s_2} \right)\Bigg \vert_
{h = 0,\, s_1=0,\,s_2=0}
=0 \,\, ,
\ee
we obtain 
\bea
\mu_H^2 + \lambda_H v^2 = 0\,\, .
\label{model9}
\eea
The mass matrix of the scalar part of the present model is constructed by 
evaluating the second order derivative terms of the scalar potential namely
$\frac {\partial^2 V} {\partial h^2}$,
$\frac {\partial^2 V} {\partial s_1^2}$,
$\frac {\partial^2 V} {\partial s_2^2}$,
$\frac {\partial^2 V} { { \partial h}{\partial s_1}}$,
$\frac {\partial^2 V} { { \partial h}{\partial s_2}}$,
$\frac {\partial^2 V} { { \partial s_2}{\partial s_1}}$
at $h=s_1=s_2=0$, as
\bea
{\cal M}^{2}_{\rm scalar} &=& \left(\begin{array}{ccc}
2\lambda_H v^2 & 0 & 0 \\
0 & \mu_{S_1}^2 + \lambda_{H S_1}v^2 & 0  \\
0 & 0 & \mu_{S_2}^2 + \lambda_{H S_2}v^2
\end{array}
\right)\,\, ,
\label{model10}
\eea
where 
\bea
m^2_h = \mu^2_h + 3 \lambda_h v^2 = 2 \lambda_h v^2 \,\, ,
\label{model11}
\eea

\bea
m^2_{s_1} = \mu^2_{S_1} + \lambda_{H S_1} v^2 \,\, ,
\label{model12}
\eea

\bea
m^2_{s_2} = \mu^2_{S_2} + \lambda_{H S_2} v^2 \,\, ,
\label{model13}
\eea

\bea 
m^2_{h s_1} = m^2_{h s_2} = m^2 {s_1 s_2} \,\, .
\label{model14}
\eea

It is to be noted that the mass matrix is diagonal as there is no mixing term 
between the scalars ($h, s_1 $ and $s_2$).

\section{Electroweak phase transitions and Gravitational Waves production in 
two component FIMP DM model}
In this section we discuss the first-order electroweak phase transition 
(FOEWPT) and we would also like to explore the generation of the Gravitational 
Waves from FOEWPT in context of the considered two component FIMP dark matter 
model.

\subsection{Effective potential at finite temperature}
In order to study the electroweak phase transition (EWPT) in the present extended SM (discussed in Section 2), 
the finite temperature correction is required to be considered with 
the tree level scalar potential. The effective potential at finite temperature 
$T$ is given as \cite{Wainwright:2012}
\bea
V_{\rm {eff}}=V_{\rm tree-level}^{T=0}+V_{\rm 1-loop}^{T=0}+V_{\rm 1-loop}^
{T\neq0},
\label{gw1}
\eea
where $V_{\rm tree-level}^{T=0}, V_{\rm 1-loop}^{T=0}$ and 
$V_{\rm 1-loop}^{T\neq0}$ define the tree level potential at $T=0$, 
the Coleman - 
Weinberg one loop corrected potential at $T=0$ and the same at finite 
temperaure ($T\ne0$) respectively. By replacing $s_1$ and $s_2$ fields with their 
corresponding classical fields $v_1$ (VEV of $S_1$) and $v_2$ (VEV of $S_2$), 
we can express the tree level potential ($V_{\rm tree-level}^{T=0}$) as
 
\bea
V_{\rm tree-level}^{T=0}&=& \frac{\mu_H^2} {2} v^2 + \frac{\lambda_H} {4} 
v^4 +
\frac{\mu_{S_1}^2} {2} v_1^2 + \nonumber \\
&& \frac{\lambda_{S_1}} {4} v_1^4 +
\frac{\mu_{S_2}^2} {2} v_2^2 + \frac{\lambda_{S_2}} {4} v_2^4 + \nonumber\\
&& \frac{\lambda_{HS_1}} {2} v^2 v_1^2 +
\frac{\lambda_{HS_2}}{2} v^2 v_2^2 +
\lambda_{S_1 S_2} v_1^2 v_2^2 \,\, .
\label{gw2}
\eea
The Coleman - Weinberg one loop corrections of the effective potential 
at zero temperature $V_{\rm 1-loop}^{T=0}$ is written as 
\cite{Wainwright:2012,{coleman}} 
\bea
V_{\rm 1-loop}^{T=0}=\pm \dfrac{1}{64\pi^2} \sum_{i} n_i m_i^4 \left[ \log\dfrac{m_i^2}{Q^2}-C_i\right],
\label{gw2a}
\eea
where $+(-)$ indicate bosons (fermions) and the summation is over $i$ particles ($\equiv h, s_1, s_2, W^{\pm}, z, t$). 
In Eq. (\ref{gw2a}) the number of degrees of freedom is represented by 
$n_i$. The degrees of freedom for the gauge bosons ($W^{\pm}, Z$), fermions 
($t$), Higgs bosons ($h$) and scalars ($s_1,s_2$) are $n_{W^{\pm}} = 6, 
n_Z = 3, n_t =12, n_h = 1$ and $n_{s_1,s_2} = 1$. The field dependent 
square masses of the gauge bosons, fermions and scalars in terms of their 
corresponding classical fields $v$ (VEV of $H$), $v_1$ (VEV of $S_1$) and 
$v_2$ (VEV of $S_2$) at zero temperature can be written as 
 
\bea
m^2_t = \displaystyle\frac {1} {2} y^2_t v^2 \,\, ,
\label{gw3}
\eea

\bea
m^2_W = \displaystyle\frac {1} {4} g^2 v^2 \,\, ,
\label{gw4}
\eea

\bea
m^2_Z = \displaystyle\frac {1} {4} (g^2+g'^2) v^2 \,\, ,
\label{gw5}
\eea

\bea
m^2_h = \mu^2_H + 3 \lambda_H v^2 +  \lambda_{H S_1} v^2_1 + \lambda_{H S_2} v^2_2\,\, ,
\label{gw6}
\eea

\bea
m^2_{s_1} = \mu^2_{s_1} + \lambda_{H S_1} v^2 + 3 \lambda_{S_1} v^2_1 + 2 \lambda_{S_1 S_2} v^2_2\,\, ,
\label{gw7}
\eea

\bea
m^2_{s_2} = \mu^2_{s_2} + \lambda_{H S_2} v^2 + 3 \lambda_{S_2} v^2_2 + 2 \lambda_{S_1 S_2} v^2_1\,\, ,
\label{gw8}
\eea

\bea
m^2_{G^+} = 2 \mu^2_H + 2 \lambda_H v^2 + 2 \lambda_{H S_1} v^2_1 + 2 \lambda_{H S_2} v^2_2 \,\, ,
\label{gw9}
\eea

\bea
m^2_{G_0} = \mu^2_H + \lambda_H v^2 + \lambda_{H S_1} v^2_1 +  \lambda_{H S_2} v^2_2 \,\, ,
\label{gw10}
\eea
where $y_t$, $g$ and $g'$ correspond to the top Yukawa coupling, SU(2)$_L$ and U(1)$_Y$ gauge SM couplings respectively. 
It is to be noted that for $v_1 = v_2 = 0$ the Eqs. (\ref{gw6}) - (\ref{gw8}) will be similar to the Eqs. (\ref{model11}) 
- (\ref{model13}) mentioned in Section 3. Here we impose Landau gauge ($\xi = 0$), where at a finite temperature ($T\ne0$) 
the Goldstone bosons acquire masses but for zero temperature they are massless and also the ghost contribution does 
not exist here \cite{Basler:2016obg}. 
For further calculations, we consider the values of the renormalisable scale, 
$Q$ (mentioned in Eq. (\ref{gw3})) as $\sim$ 246.22 GeV. However, one can vary $Q$ as the selection of $Q$ is not unique. 
In Eq. (\ref{gw2a}) $C_i$ indicates a numerical constant, which depends on the renormalisation scheme. For the gauge bosons ($W^{\pm}, Z$) and for the other particles considered in this model ($h, s_1, s_2,t$), the 
numerical constant can be taken as $C_{W, Z} = 5/6$ and $C_{h, s_1, s_2, t} = 3/2$. The one loop corrected effective 
potential at $T \ne 0$ takes the form

\bea
V_{1-{\rm loop}}^{T\neq0}=\dfrac{T^2}{2\pi^2} \sum_{i} n_i J_{\pm}\left[\dfrac{m_i^2}{T^2}\right],
\label{gw11}
\eea

where the function $J_{\pm}$ is given as 
\bea
J_{\pm}\left(\dfrac{m_i^2}{T^2}\right)=\pm \int_0^{\infty} dy 
\hspace{1mm} y^2 \log\left(1\mp e^{-\sqrt{y^2+\dfrac{m_i^2}{T^2}}}\right).
\label{gw12}
\eea
We add the temperature corrected terms to the boson mass by including the Daisy resummation procedure in the one 
loop corrected effective potential $V_{1-{\rm loop}}^{T\neq0}$ \cite{Arnold} 
at the finite temperature. Therefore the temperature dependent 
thermal mass can be expressed as $\mu^{\prime 2}_{H}(T)=\mu_H^2 + c_1 T^2$, 
$\mu^{\prime 2}_{S_2}(T)=\mu_{S_2}^2+c_2 T^2$ and 
$\mu^{\prime 2}_{S_3} (T)=\mu_{S_3}^2+c_3 T^2$, where

\bea
c_1=\dfrac{6\lambda_H+2\lambda_{H S_1}}{12}+\dfrac{3g^2+{g^{\prime}}^2}{16}+\dfrac{1}{6}\lambda_{H S_2}
+\dfrac{y_t^2}{4}\,\,,
\label{gw13}
\eea

\bea
c_2= \dfrac{1}{8}\lambda_{S_1}+\dfrac{1}{6}\lambda_{H S_1}+\dfrac{1}{6}\lambda_{S_1 S_2}\,\,,
\label{gw14}
\eea

\bea
c_3=\dfrac{1}{8}\lambda_{S_2}+\dfrac{1}{6}\lambda_{H S_2}+\dfrac{1}{6}\lambda_{S_1 S_2}\,\,.
\label{gw15}
\eea

For our calculation the CosmoTransition package \cite{Wainwright:2012} has been used to compute the correction to the tree-level potential 
at finite temperature.

\section{Gravitational Waves production from FOEWPT}
The onset of bubble nucleation when the system attains a vacuum state at a finite temperature is central to the occurence 
of first-order phase transition. This temperature when the bubble nucleation sets in is generally known as the 
nucleation temperature. All possible shapes of bubbles having 
different internal pressure and surface tensions can be produced in this process. The bubbles having size smaller than 
the critical size (the size at which the bubbles can just avoid any collapse) tend to collapse, whereas the 
larger ones tend to expand after achieving the criticality. The bubbles lose their spherical symmetry when they collide 
with each other. This process plays a major role to initiate the phase transitions and the emissions of 
Gravitational Waves.

We can write the bubble nucleation rate per unit volume at a particular temperature ($T$) as \cite{Linde:1983}
\bea
\Gamma=\Gamma_0\left(T \right) e^{-S_3\left(T \right)/T}\,\, ,
\label{pt1}
\eea
where $\Gamma_0\left(T \right) \propto T^4$. In the above, $S_3 (T)$ signifies the Euclidean action of the 
critical bubble, which is given 
as 
\bea
S_{3}=4\pi \int dr \hspace{1mm}r^{2} \left[ \dfrac{1}{2} \left(\partial_{r} \vec{\phi} \right)^2 +V_{\rm eff}\right],
\label{pt2}
\eea
where $V_{\rm eff}$ denotes the effective finite temperature potential (Eq. (\ref{gw1}). The bubble nucleation occurs 
when the nucleation temperature 
($T_n$) satisfies the condition $\dfrac{S_3 (T_n)}{T_n} \simeq 140$. The three 
mechanisms via which the Gravitational Waves are generated from FOEWPT are bubble collisions \cite{Kosowsky:1992a}-\cite{Caprini:2008}, sound wave \cite{Hindmarsh:2014}-\cite{Hindmarsh:2015} and turbulance 
effect in the plasma \cite{Caprini:2006}-\cite{Caprini:2009}. By considering the contributions of the three above mentioned mechanisms, the total GW intensity 
($\Omega_{\rm GW} {\rm h}^2$) as a function of frequency can be written as \cite{Kosowsky:1992a}-\cite{Caprini:2009}
\bea
\Omega_{\text{GW}}{\rm h}^2=\Omega_{\text{col}}{\rm h}^2+ \Omega_{\text{SW}}{\rm h}^2+ \Omega_{\text{turb}}{\rm h}^2.
\label{pt3}
\eea
where $\Omega_{\rm col} {\rm h}^2$ denotes the contribution of the bubble collisions to the total GW intensity 
(for more detail derivations see Ref \cite{col1,{col2}}), which can be expressed as

\bea
\Omega_{\text{col}}{\rm h}^2=1.67\times 10^{-5} \left(\dfrac{\beta}{H} \right) ^{-2} \dfrac{0.11 v_{w}^3}{0.42+v_{w}^2} \left(\dfrac{\kappa \alpha}{1+\alpha}\right)^2 \left(\dfrac{g_*}{100}\right)^{-\frac{1}{3}}\dfrac{3.8 \left(\dfrac{f}{f_{col}}\right)^{2.8}}
{1+2.8 \left(\dfrac{f}{f_{\text{col}}}\right)^{3.8}}\,\,\,,
\label{pt4}
\eea
with the parameter $\beta$ 
\bea
\beta=\left[H T \dfrac{d}{dT}\left( \dfrac{S_3}{T}\right) \right]\bigg|_{T_n}.
\label{pt5}
\eea
In the above equation (Eq. (\ref{pt5})), $H = H_n$, the Hubble parameter 
at nucleation temperature $T_n$. The bubble wall velocity ($v_w$) is generally estimated as \cite{Kamionkowski:1994, {Steinhardt:1982}}

\bea
v_w=\dfrac{1/\sqrt{3}+\sqrt{\alpha^2+2\alpha/3}}{1+\alpha}.
\label{pt6}
\eea
The choice of the bubble wall velocity varies from literature to literature \cite{Vaskonen:2016yiu,{Madge:2018gfl}}. For simplicity in some literatures $v_w$ is considered to be 1 \cite{Shajiee:2018jdq, {Mohamadnejad:2019vzg}, {Chala:2019rfk}}. In our work we use Eq. (\ref{pt6}) for the calculation of bubble wall velocity.

The fraction of latent heat deposited in a thin shell denoted by $\kappa$ in Eq. (\ref{pt4}) is given as
\bea
\kappa=1-\dfrac{\alpha_{\infty}}{\alpha},
\label{pt7}
\eea
with \cite{Shajiee:2018jdq,{Caprini:2015zlo}}
\bea
\alpha_{\infty}=\dfrac{30}{24\pi^2 g_{*}} \left(\dfrac{v_n}{T_n} \right)^2 \left[6 \left( \dfrac{m_W}{v}\right)^2 +3\left( \dfrac{m_Z}{v}\right)^2 +6\left( \dfrac{m_{t}}{v}\right)^2\right].
\label{pt8}
\eea
where $v_n$ indicates the VEV of Higgs at the nucleation temperature ($T_n$). The masses of $W^{\pm}, Z$ and top quarks 
are represented as $m_W, m_Z$ and $m_t$ repectively in the above. The parameter $\alpha$ is defined as the ratio of the vacuum energy 
density ($\rho_{\rm vac}$) released by EWPT to the background energy density ($\rho^*_{\rm rad}$) at $T_n$. Therefore 
the parameter $\alpha$ can be expressed as
\bea
\alpha=\left[\dfrac{\rho_{\text{vac}}}{\rho^*_{\text{rad}}}\right]\bigg|_{T_n}.
\label{pt9}
\eea
where $\rho_{\rm vac}$ and $\rho^*_{\rm rad}$ can be written as 
\bea
\rho_{\text{vac}}=\left[\left(V_{\text{eff}}^{\text{high}}-T\dfrac{dV_{\text{eff}}^{\text{high}}}{dT} \right)-\left(V_{\text{eff}}^{\text{low}}-T\dfrac{dV_{\text{eff}}^{\text{low}}}{dT} \right)\right],
\label{pt10}
\eea
and
\bea
\rho^*_{\text{rad}}=\dfrac{g_* \pi^2 T_n^4}{30}.
\label{pt11}
\eea
It is to be mentioned that one may calculate $\rho_{\rm vac}$ by using the trace anomaly \cite{Hindmarsh:2015}, where we need to include an additional factor 
$\dfrac{1}{4}$ with the term $T\dfrac{dV}{dT}$ (Eq. (\ref{pt10})).

The peak frequency produced by the bubble collisions (the parameter $f_{\rm col}$ (Eq. (\ref{pt4})) can be expressed as 
\bea
f_{\text{col}}=16.5\times10^{-6}\hspace{1mm} \text{Hz} \left( \dfrac{0.62}{v_{w}^2-0.1 v_w+1.8}\right)\left(\dfrac{\beta}{H} \right) \left(\dfrac{T_n}{100 \hspace{1mm} \text{GeV}} \right) \left(\dfrac{g_*}{100}\right)^{\frac{1}{6}}.
\label{pt12}
\eea
In Eq. (\ref{pt3}), $\Omega_{\rm SW} {\rm h}^2$ is the sound wave component of the GW intensity which takes the form 
\bea
\Omega_{\text{SW}}{\rm h}^2=2.65\times 10^{-6} \left(\dfrac{\beta}{H} \right) ^{-1} v_{w} \left(\dfrac{\kappa_{v} \alpha}
{1+\alpha}\right)^2 \left(\dfrac{g_*}{100}\right)^{-\frac{1}{3}}\left(\dfrac{f}{f_{\text{SW}}}\right)^{3} \left[\dfrac{7}
{4+3 \left(\dfrac{f}{f_{\text{SW}}}\right)^{2}}\right]^{\frac{7}{2}}.
\label{pt13}
\eea
In the above equation, $\kappa_v$ defines the fraction of latent heat transformed into the bulk motion of the fluid which 
is given as 
\bea
\kappa_v=\dfrac{\alpha_{\infty}}{\alpha}\left[ \dfrac{\alpha_{\infty}}{0.73+0.083\sqrt{\alpha_{\infty}}+\alpha_{\infty}}\right].
\label{pt14}
\eea
In Eq. (\ref{pt13}) the peak frequency $f_{\rm SW}$ produced by the sound wave mechanisms has the following form 
\bea
f_{\text{SW}}=1.9\times10^{-5}\hspace{1mm} \text{Hz} \left( \dfrac{1}{v_{w}}\right)\left(\dfrac{\beta}{H} \right) \left(\dfrac{T_n}{100 \hspace{1mm} \text{GeV}} \right) \left(\dfrac{g_*}{100}\right)^{\frac{1}{6}}.
\label{pt15}
\eea
In order to check whether the contribution of sound wave component to the total GW intensity is significant or not, 
it is useful to estimate a factor $\dfrac{HR_*}{\bar{U}_f}$, known as suppression factor $\dfrac{HR_*}{\bar{U}_f}$, 
where $\bar{U}_f$ and $R_*$ are the root-mean-square (RMS) fluid velocity and 
the mean bubble separation respectively \cite{Caprini:2015zlo,{Ellis:2018mja},{Ellis:2019oqb}}. For a given model if the suppression factor $\left (\dfrac{HR_*}{\bar{U}_f} 
\right)$ turns 
out to be $<$ 1, then  we need to include this factor to the sound wave component of the GW intensity as it is an 
overestimate to the GW signal. On the other hand if $\dfrac{HR_*}{\bar{U}_f}$ $>$ 1 , then the sound wave survives more 
than Hubble time.

The third mechanism which takes part in the production of GWs is the turbulance component 
($\Omega_{\text{turb}}{\rm h}^2$) arising out of the turbulance in plasma
\bea
\Omega_{\text{turb}}{\rm h}^2=3.35\times 10^{-4} \left(\dfrac{\beta}{H} \right) ^{-1} v_{w} \left(\dfrac{\epsilon \kappa_v \alpha}{1+\alpha}\right)^{\frac{3}{2}} \left(\dfrac{g_*}{100}\right)^{-\frac{1}{3}} \dfrac{\left(\dfrac{f}{f_{\text{turb}}}\right)^{3}\left( 1+\dfrac{f}{f_{\text{turb}}}\right)^{-\frac{11}{3}}}{\left(1+\dfrac{8\pi f}{h_{*}}\right)},
\label{pt16}
\eea
where $\epsilon = 0.1$ and the peak frequency $f_{\rm turb}$ is 
\bea
f_{\text{turb}}=2.7\times10^{-5}\hspace{1mm} \text{Hz} \left( \dfrac{1}{v_{w}}\right)\left(\dfrac{\beta}{H} \right) \left(\dfrac{T_n}{100 \hspace{1mm} \text{GeV}} \right) \left(\dfrac{g_*}{100}\right)^{\frac{1}{6}}.
\label{pt17}
\eea
In Eq. (\ref{pt16}), the parameter $h_*$ takes the form 
\bea
h_{*}=16.5\times10^{-6}\hspace{1mm} \text{Hz} \left(\dfrac{T_n}{100 \hspace{1mm} \text{GeV}} \right) \left(\dfrac{g_*}{100}\right)^{\frac{1}{6}}.
\label{pt18}
\eea
The GW intensity can now be computed using Eqs. (\ref{pt3}) - (\ref{pt18}).
\section{Calculations and results}
In this section, we calculate the intensity of Gravitational Wave signals
produced from the first-order electroweak
phase transition (FOEWPT) in a two component FIMP dark matter model
considered in this work and the calculated signals
are compared with the sensitivity
curves of such GWs at different proposed future detectors such as
ALIA, BBO, DECIGO, aLIGO, aLIGO+ and LISA. In order to
compute the GW intensities we choose three sets of benchmark
points (BPs) from the allowed model parameter space. These three BPs are chosen
in such a way that they satisfy various bounds and constraints on
the model parameters from both theoretical considerations
(vacuum stability \cite{vacuum}, perturbativity \cite{vacuum}-\cite{lee} etc.)
and experimental observations, e.g. the PLANCK observational
results for the dark matter relic density, the collider bounds
and the upper limit on the dark matter self interaction
cross-section.
All the necessary constraints mentioned above and the relic density
calculations of the present model have been worked out
in detail in an earlier work \cite{madhu}. In Ref. \cite{madhu} the authors
have shown that the two real singlet scalars constitute a viable two component dark matter
candidate. The considered BPs are tabulated in Table 1. In addition
to the model parameters we have
also furnished in Table 1, the values of the dark matter relic densities 
corresponding to each of
the three BPs.

\begin{table}[H]
\centering
\footnotesize
\begin{tabular}{|l|c|c|c|c|c|c|c|c|c|c|r|}
\hline
BP & $m_h$ & $m_{s_{1}}$ & $m_{s_2}$ & $\lambda_{H S_1}$ & $\lambda_{H S_2}$ & $\lambda_{S_1}$ & $\lambda_{S_2}$ &
$\lambda_{S_1 S_2}$ & $\Omega_{\text{DM}}{\rm h}^2$\\
 & in GeV & in GeV & in GeV &&&&&& \\
 \hline
1 & 125.5 & $10 \times 10^{-6}$ & $5 \times 10^{-6}$ & $2 \times 10^{-9}$ & $1 \times 10^{-9}$ & 0.89 & 1.36 & 0.71 & 0.1217 \\
 \hline
2 & 125.5 & 0.01 & 0.005 & $6.36 \times 10^{-11}$ & $2.67 \times 10^{-11}$ & 1.68 & 1.97 & 1.90 & 0.119 \\
 \hline
3 & 125.5 & $10 \times 10^{-6}$ & $5 \times 10^{-6}$ & $2 \times 10^{-9}$ & $1 \times 10^{-9}$ & 2.0 & 0.1 & 0.01 & 0.1217 \\
\hline
\end{tabular}
\caption{The chosen three benchmark points (BPs) are tabulated. We use the parameter values of these chosen BPs to calculate the GW intensity
produced from a two component FIMP dark matter model. The relic
density values corresponding to each of the BPs are furnished in this table.}
\label{t1}
\end{table}

A discussion is in order. The choice of the self couplings in Table \ref{t1}
of the two components $s_1$, $s_2$ of the FIMP dark matter are made as follows.
In Ref. \cite{madhu}, the calculations are made for similar model (Fig. 12 of Ref.
\cite{madhu}) by varying the self couplings from 0 to $\frac {2\pi} {3}$. In this
work we found that smaller choices of self couplings (although allowed)
does not yield significant GW intensity. The values of self couplings in
Table 1 are therefore chosen within the allowed limits so as to obtain
strong first-order phase transitions and the subsequent production of GWs.

The thermal parameters which play a major role for GW emissions are
the time-scale of the phase transition ($1/\beta$),
the strength of the first-order phase transition ($\alpha$), the bubble
wall velocity ($v_w$), the nucleation temperature ($T_n$) and
VEV of Higgs ($v_n$) at $T_n$. To estimate the GW intensity, firstly
we need to compute the above mentioned thermal parameters
and for that we consider a finite temperature effective potential
which is obtained by adding the thermal
correction terms with the tree level potential (Eq. (\ref{gw2})).
The calculations related to the thermal parameters
have been done using Cosmotransition package \cite{Wainwright:2012}. The calculated
values of the phase transition parameters
such as $v_n, v_c, T_c, v_c/T_c, T_n, \alpha, \beta/H$ are
furnsihed in Table 2. All the chosen BPs satisfy the strong
first-order phase transition (SFOPT) condition $v_c/T_c > 1$
($v_c$ is the VEV of Higgs at critical temperature
$T_c$) which is shown in Table 2. Here we note again that calculations made with choice of low values 
of dark matter self couplings ($ \lesssim 10^{-2}$) do not
yield results feasible for SFOEWPT and eventual production of Gravitational
Waves. Therefore, in this scenario the DM self coupling influences the
GW production. It is to be noted that the phase
transition occurs in between the critical temerature
and the temperature at the present epoch. Therefore the nucleation
temperature ($T_n$) is always smaller
than the critical temperature ($T_c$) and this is true for all the
considered BPs (Table 2). For all the calculations
the renormalisable scale ($Q$) is kept fixed at 246.22 GeV.
We use Eqs. (\ref{pt3})-(\ref{pt18}) to compute the intensity of the
GW. As mentioned earlier, in order to estimate the contribution
of the sound wave component to the total GW intensity, the suppression
factor $\dfrac{HR_*}{\bar{U}_f}$ is computed (following Refs. \cite{Caprini:2015zlo,{Ellis:2018mja},{Ellis:2019oqb}}) and we obtain that
$\dfrac{HR_{*}}{\bar{U}_f} < 1$ for all the three chosen BPs.

\begin{table}[H]
\centering
\footnotesize
\begin{tabular}{|l|c|c|c|c|c|c|c|c|c|r|}
\hline
BP & $v_n$ & $v_c$ & $T_c$ & $\dfrac{v_c}{T_c}$ & $T_{n}$ & $\alpha$ & $\dfrac{\beta}{H}$ &
$\dfrac{HR_*}{\bar{U}_f}$\\
 & in GeV & in GeV & in GeV &&&&&\\
 \hline
1 & 277.07 & 256.51 & 94.15 & 2.72 & 86.62 & 0.30 & 1916.1 & 0.07\\
\hline
2 & 275.98 & 260.30 & 98.06 & 2.65 & 91.15 & 0.29 & 3168.35 & 0.228\\
\hline
3 & 300.51 & 288.2 & 117.2 & 2.46 & 103.47 & 0.40 & 941.67 & 0.228\\
\hline
\end{tabular}
\caption{The values of the thermal parameters used for computing the GW intensity for each of the corresponding chosen
BPs. See text for details.}\label{t2}
\end{table}

The strong first-order electroweak phase transition (SFOEWPT) properties for BP 1 are shown in Figure 1 (a and b).
In Figure 1(a) we have plotted the
phase structure of the model as a function of the temperature for the choice, BP 1. From the Figure 1(a) it is observed that the
SFOEWPT occurs at the nucleation temperature $T_n$ = 86.62 GeV and at this temperature a potential separation between a
high (indicated as blue line) and low (indicated as orange line) phase appears. We also study the phase transition
properties of other two BP points (BP2, BP3) and we found that the nature of the plots are similar to what is shown in
Figure 1(a) (BP1). The parameter $\beta$ (Eq. (\ref{pt5})) can be estimated from the slope of the plot $S_3/T$ vs $T$
(Figure 1(b)) around the nucleation temperature ($T_n$). This is seen to satisfy the condition $S_3/T_n$ = 140. For the
demonstating purpose we show the variation of the parameter
$S_3/T$ with the temperature $T$ for BP1 in Figure 1(b).

In Figure 2 the calculated GW intensities for three BPs have been plotted as a function of frequency and
we make a comparison with the estimated detectability \footnote[1]{We consider the power-law-integrated sensitivity approach \cite{Thrane:2013oya,Dev:2019njv}. For
an alternate method see \cite{Moore:2015,{Alanne:2019bsm},{Schmitz:2020syl}}.} of such GWs at the future generation ground based telescopes
(aLIGO and aLIGO+) and space based telescopes (ALIA, BBO, DECIGO and LISA). For BP1, BP2 and BP3 the GW intensities acquire
peaks at the frequencies $7 \times 10^{-3}$ Hz, $1.2 \times 10^{-2}$ Hz and $4 \times 10^{-3}$ Hz respectively. It is
observed from Figure 2 that the GW intensity for all the BPs (BP1 - BP3) fall within the sensitivity curves of ALIA, BBO
and DECIGO. From Table 2 we obtain the highest value of the parameter $\alpha$ and the lowest value of the parameter
$\dfrac{\beta}{H}$ for BP3 compared to the other two benchmark points (BP1, BP2) which reflects the fact that the GW
intensity is higher for BP3 and it is also evident from Figure 2. Therefore, it appears that the principal dependence of
GW intensity is on the parameters $\alpha$ and $\dfrac{\beta}{H}$.

\begin{figure}[H]
\centering
\subfigure[]
{\includegraphics[width=7cm,height=5.5cm]{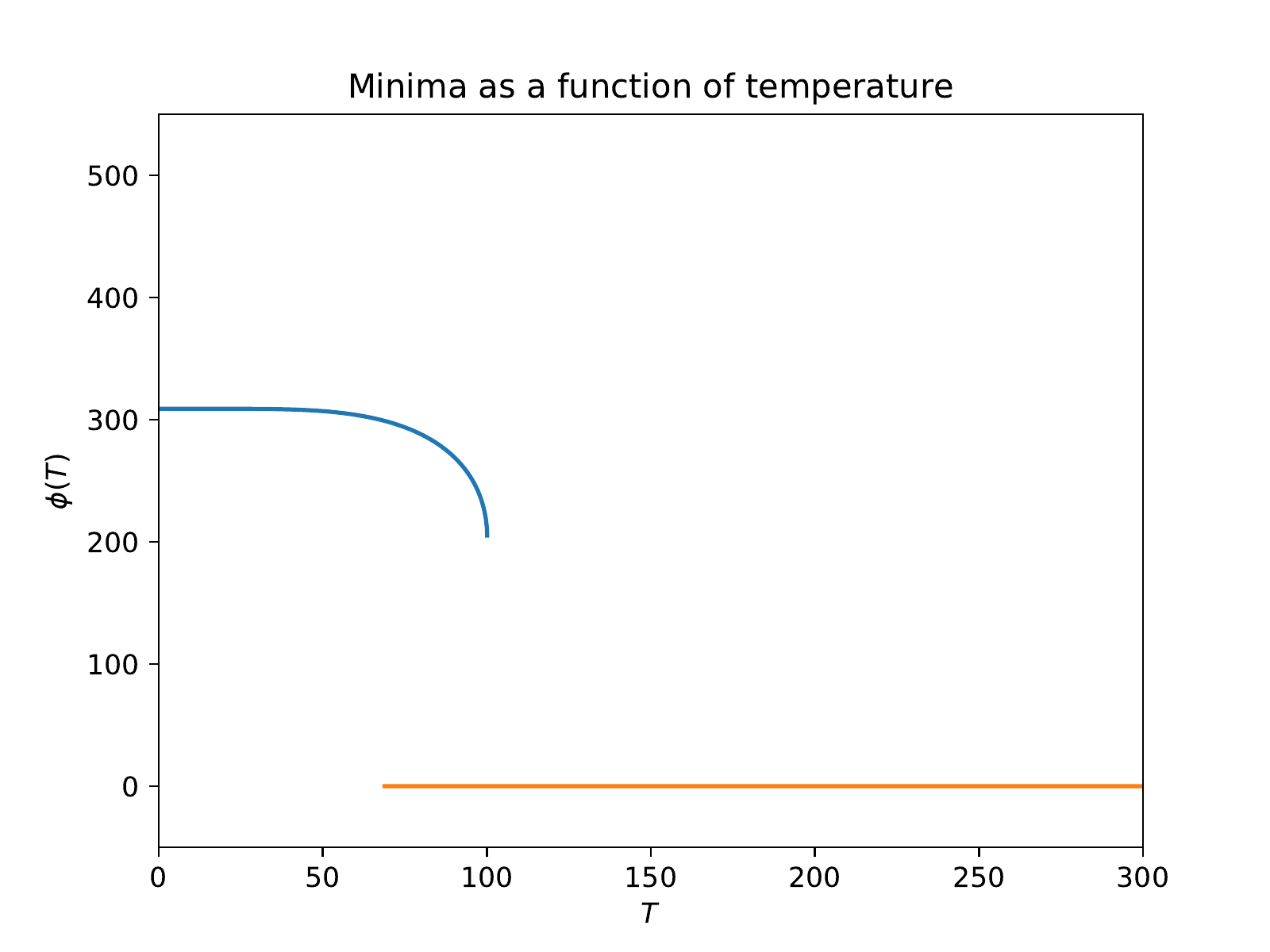}}
\subfigure[]
{\includegraphics[width=7cm,height=5.5cm]{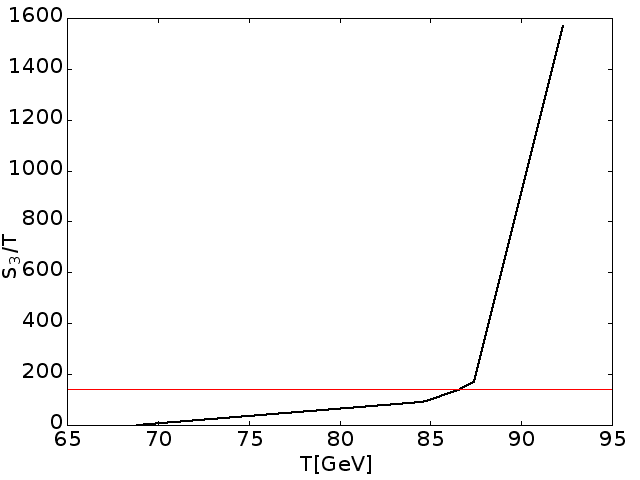}}
\caption{Variations of the phase transition properties with temperature for BP1. In (a) we show how the 
phase factor ($\phi (T)$) varies with the temperaure and it creates a potential separation at $T_n$ = 86.62 GeV. Figure 
1(b) gives the variation of the parameter $S_3/T$ with $T$ and the red solid line indicates that nucleation occurs for 
the condition $S_3/T$ =140. See text for details.}
\label{fig:1}
\end{figure}

\begin{figure}[H]
\centering
\includegraphics[width=9cm,height=8cm]{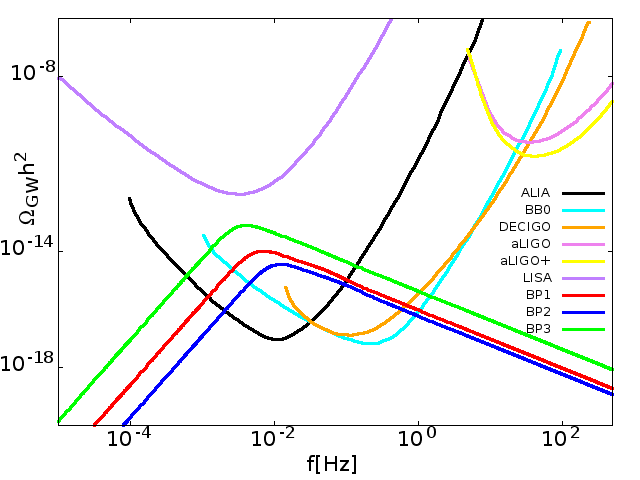}
\caption{The variations of the GW intensity with frequency for each of the three BPs and their comparisons 
with the sensitivity curves of the detectors namely aLIGO, aLIGO+, ALIA, BBO, DECIGO and LISA. 
See text for details.}
\label{fig:2}
\end{figure}

\section{Summary and Conclusions}
In this work, we have considered the emission of Gravitational Waves
from first-order electroweak phase transitions involving a two component
FIMP dark matter and the detectabilities of these Gravitational Waves
by some proposed space based detectors in future.
We consider a two
component dark matter model where the scalar sector of Standard
Model is extended by introducing two real scalars. The added scalars
are singlets under SM gauge group SU(2)$_{\rm L} \times $ U(1)$_{\rm Y}$
and constitute the two components of the present dark matter model.
Both the components of this two component dark matter model are
produced via ``freeze-in" mechanism whereas their abundances grow from
negligible amount towards the equilibrium value. These are 
Feebly Interacting Massive Particle or FIMP and in the present scenario
their masses are within the range of few keV to few MeV.
A discrete $Z_2 \times Z'_2$ symmetry is imposed on the two scalars
to ensure their stability and they do not acquire any VEV at SSB.
The extensive study of the dark matter phenomenology
of the considered model as well as the theoretical and
experimental constraints on the model parameters of the
model have been discussed by one of the authors in a previous work \cite{madhu}.
We explore the possibility of first-order electroweak phase
transition or EWPT with
this model and subsequent productions of GWs. From the allowed
parameter space we choose three benchmark points for both
the analytical and numerical calculations of the GW intensity.
Finite temperature corrections of the tree-level potential
have been introduced for the calculations of the GW signals. For the
model chosen parameters we compute the intensities of Gravitational
Waves from the first-order EWPT initiated by the present extended SM.
We compare our results with the projected sensitivities of future
space based (ALIA, BBO, DECIGO amd LISA) as well as
ground based (aLIGO and aLIGO+) GW detectors that would detect such
primordial GWs. From our calculations it is observed
that the peak values of such GWs lie within the detectable range of
the future detectors such as ALIA, BBO and DECIGO. We also find that
in the present dark matter model though the DM phenomenology
(e.g. relic density calculations) can be addressed by
considering very small Higgs portal couplings in FIMP mechanism,
the production of the GW signals during the
DM assisted EWPT is mostly governed by the couplings of the dark matter
self interactions. Thus our
two component FIMP dark matter model is in addition to a viable model
for particle dark matter, it can cause GW productions via first-order
electroweak phase transitions.

{\bf Acknowledgements}: The authors would like to thank D. Majumdar for his useful comments and
suggestions. The authors would also like to thank B. Banerjee for helping in the CosmoTransition package modification. 
One of the authors (MP) thanks the DST-INSPIRE fellowship grant (DST/INSPIRE/FELLOWSHIP/[IF160004]) by DST. Govt. of India.

\end{document}